\documentclass{aa}
\usepackage[varg]{txfonts}
\usepackage{multirow}
\usepackage{natbib}
\usepackage{graphicx}

\begin{document}

\title{Tidal capture and repeating partial tidal disruption events of giant stars}
\author{Di Wang\inst{\ref{inst1}} \and Fa-Yin Wang\inst{\ref{inst1},\ref{inst2}}}
\institute{School of Astronomy and Space Science, Nanjing University, Nanjing 210093, People’s Republic of China\\
\email{wangdi@nju.edu.cn}
\\
\email{fayinwang@nju.edu.cn}\label{inst1}
\and 
Key Laboratory of Modern Astronomy and Astrophysics (Nanjing University), Ministry of Education, Nanjing 210093, People's Republic of China\label{inst2}
}
\abstract
{
When an object is scattered near a supermassive black hole (SMBH), tidal oscillations excited within it reduce its orbital energy, leading to capture by the SMBH. This process, called tidal capture, can also occur when the object closely approaches the SMBH, resulting in a partial tidal disruption event (pTDE). Previous studies on pTDEs of main-sequence stars have shown that as the disruption intensifies, dynamical effects dominate over tidal oscillations, causing the remnant material to acquire a kick velocity instead of being captured by the SMBH. In this work, we performed hydrodynamic numerical simulations of pTDEs involving giant stars. We find that for weaker disruptions, the dynamical behavior of the remnant material resembles that of main-sequence stars. However, as the disruptions deepen, the remnant material transitions from gaining energy to losing energy, leading to capture by the SMBH. This behavior markedly differs from that of main-sequence stars, demonstrating that the presence of a compact core significantly influences the dynamical processes in pTDEs. Our simulations reveal that the energy change in the remnant material strongly correlates with asymmetric mass loss—specifically, the difference in mass outflow between the Lagrange points $L_1$ and $L_2$. This suggests that the energy change stems from asymmetric mass loss, consistent with conclusions from previous studies on main-sequence stars. However, a quantitative analysis contradicts earlier models, indicating that the dynamical model of pTDEs requires further refinement. Finally, we discuss the characteristics of repeating pTDEs produced by this process and their potential observability, as well as the implications for the long-term orbital evolution of high-eccentricity, extreme-mass-ratio inspiral systems.
}
\maketitle
\nolinenumbers

\section{Introduction}

When two unbound stars encounter one another, tidal forces excite internal oscillations that extract orbital energy, thereby forming a bound pair \citep{1975MNRAS.172P..15F,1977ApJ...213..183P}. An analogous process—tidal capture—occurs when a star passes sufficiently close to a supermassive black hole (SMBH); this can induce repeated partial tidal disruption events(rpTDEs; \citealt{2013ApJ...771L..28M,2013ApJ...762...37L,2023MNRAS.520L..38C,2023ApJ...942L..33W,2024ApJ...977...80C,2024ApJ...971L..31P,2023ApJ...948...89K}). 

Numerical simulations of these encounters reveal that, for relatively distant passages, the star loses sufficient orbital energy through tidal excitation to become bound to the SMBH \citep{2013ApJ...771L..28M,2023MNRAS.520L..38C,2024ApJ...977...80C}, which could also happen around intermediate mass black holes (IMBHs; \citealt{2023ApJ...948...89K}). Conversely, for closer passages, asymmetric mass loss dominates over tidal excitation, imparting a net recoil that leaves the star on a more unbound orbit.

Previous studies have concentrated on the partial disruption of main-sequence stars. To avoid complete disruption, the periastron distance ($r_p$) must satisfy $r_p \lesssim r_t$, i.e., $\beta = r_t/r_p \lesssim 1$. However, for very large $\beta$, \citet{2022ApJ...927L..25N} show that the star can survive because the encounter time is shorter, and the remnant core is subsequently bound to the SMBH without receiving a significant kick;  however, the convergence of these simulations has not yet been achieved.

On the other hand, the orbital energy variation during the tidal disruption event (TDE) of a giant star has not been sufficiently examined. \citet{2013ApJ...762...37L} studied a tidally disrupted giant planet that has a compact core with $\beta=0.5$–$1$. They found that the asymmetric mass-loss evolution with $\beta$ differs from that for a planet without a core, and that the change in orbital energy is linearly related to the asymmetric mass loss, i.e., the difference between the mass loss through the Lagrangian points $L_1$ and $L_2$. However, they did not study a situation with $\beta>1$, where the giant planet is still partially disrupted. Because the orbital energy variation is related to the asymmetric mass loss \citep{2013ApJ...771L..28M,2013ApJ...762...37L}, it is likely different for stars with and without a core.

We performed hydrodynamical simulations to study the orbital energy variation during the TDE of the giant star. We find that the remnant core is not kicked but is instead bound to the SMBH when $\beta\gtrsim 1$, and that the linear relation between the orbital energy variation and the asymmetric mass loss remains valid. In such a situation, the bound orbital energy due to the asymmetric mass loss is much higher than that due to tidal oscillation excitation, leading to a shorter orbital period of the remnant material.

\section{Simulation setup}

We carried out the simulations with the smoothed-particle hydrodynamics (SPH) code PHANTOM \citep{2018PASA...35...31P}. The structure of the giant was generated by evolving a $1\,M_\odot$ zero-age main-sequence model in MESA until core nitrogen ignition \citep{2011ApJS..192....3P,2013ApJS..208....4P,2015ApJS..220...15P,2018ApJS..234...34P,2019ApJS..243...10P,2023ApJS..265...15J}. From the MESA output, we extracted three profiles that differ in radius and core mass, as shown in Table \ref{tab: proflies}. 

\begin{table}[htp]
    \caption{Profiles of the giant stars adopted in the simulations.}
    \label{tab: proflies}
    \centering 
    \begin{tabular}{c c c c}
        \hline\hline  
        Profile & Radius($R_{\odot}$) & Mass($M_{\odot}$)  & Core Mass ($M_{\odot}$)\\
        \hline
        RG1 & 8.189 & 0.992 & 0.236\\
        RG2 & 4.094 & 0.995 & 0.181\\
        RG3 & 2.133 & 0.996 & 0.137\\
        \hline
    \end{tabular}
\end{table}

The compact core of the giant star was replaced by a sink particle treated as a point mass that interacts with the SPH particles only through gravitation. The radius of the core is characterized by the gravitational softening radius of the sink particle. The envelope obeys an adiabatic equation of state with adiabatic index $\gamma=5/3$, and the entire star was then relaxed to hydrostatic equilibrium using $10^{6}$ SPH particles, with adaptive smoothing lengths determined by the neighbor count \citep{2018PASA...35...31P}. The stellar structures obtained through these approximations are almost identical to those in MESA (see comparisons in Appendix \ref{sec: stellar structure comparison}).

The star was initially placed in a parabolic orbit at a distance of $3r_t$ from the $10^6 M_\odot$ SMBH, where $r_t$ is the disruption radius:
\begin{equation}
    r_t=\left(\frac{M_{BH}}{M_{\star}}\right)^{1/3}R_{\star}.
\end{equation}
Here $M_{\rm BH}$ and $M_{\star}$ denote the masses of the SMBH and the giant star, respectively, and $R_{\star}$ is the radius of the giant star. The periastron distance, $r_p=r_t/\beta,$ is set by the impact factor, $\beta$. We adopted different value of $\beta$: $0.5,0.6,0.7,0.8,0.9,1,1.5,2,3,4$, and $5$. Convergence was verified by repeating selected simulations at lower resolutions. For larger $\beta$, a higher resolution is required, resulting in a higher computation cost. The simulations with higher $\beta$ were therefore not performed. All simulations were run for $20 t_{\mathrm{dyn}}$, where $t_{\mathrm{dyn}}=\sqrt{R_{\star}^3/GM_{\star}}$ is the dynamical timescale of the star.

\section{Result}
\subsection{Orbital energy variation}

The specific energy was derived from the position and velocity of the sink particle that represents the core, which is consistent with that obtained from the remnant gas (see the discussion in Appendix \ref{sec: energy in two ways}). Figure \ref{fig: epsilon_vs_time} shows the evolution of the specific orbital energy of the remnant core under different $\beta$ for profile RG2. The orbital energy changes abruptly as the star passes through periastron, especially when $\beta\gtrsim 1$. After disruption, the energy exhibits damped oscillations and a secular drift, indicating that the stripped material continues to interact with the remnant core. 

\begin{figure}[hpt]
    \centering
    \includegraphics[width=0.5\textwidth]{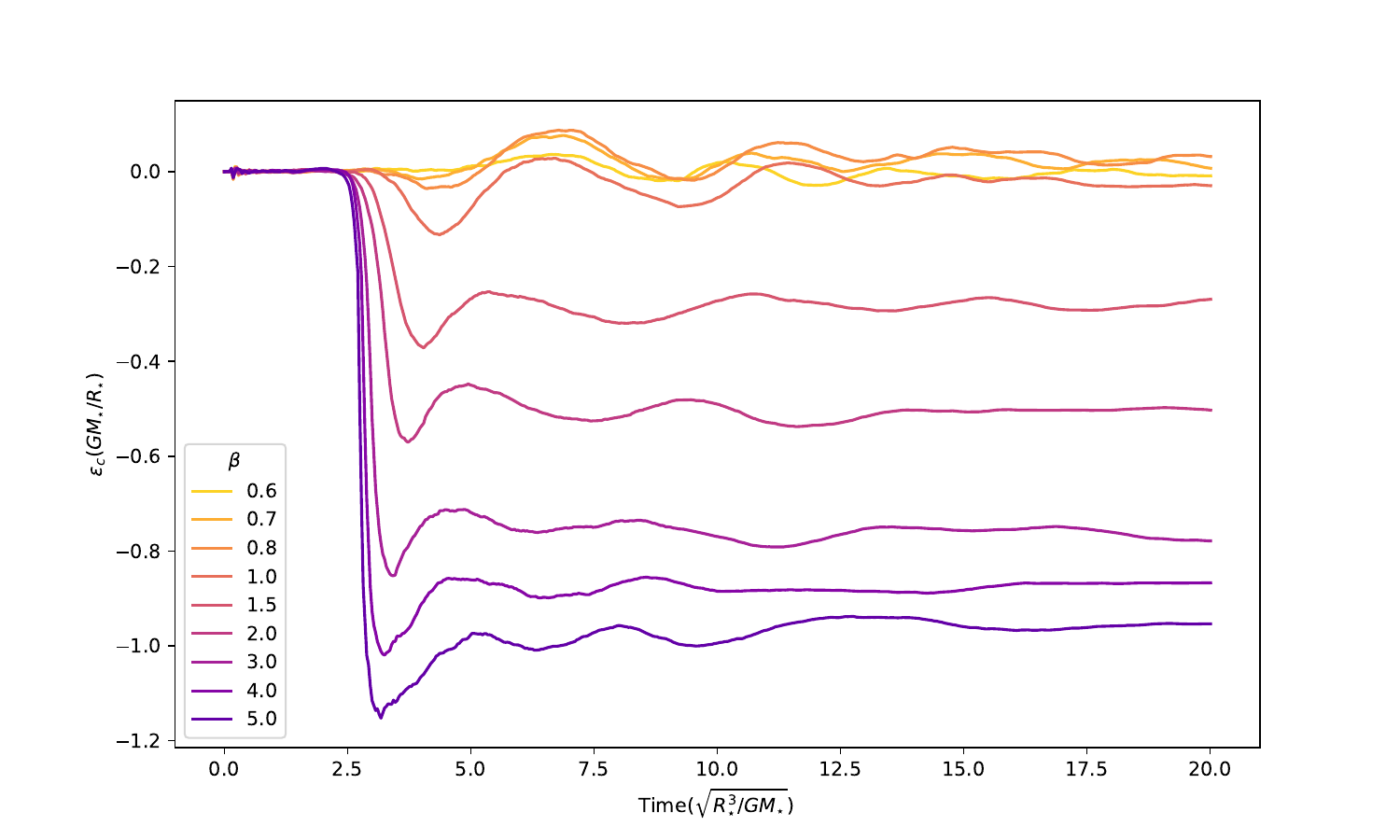}
    \caption{Specific orbital energy evolution with different $\beta$ for profile RG2. The specific orbital energy ($\epsilon_c$) and time are normalized to $GM_{\star}/R_{\star}$ and $t_{\mathrm{dyn}}=\sqrt{R_{\star}^3/GM_{\star}}$, respectively. Curves of different colors represent simulations with different $\beta$. \label{fig: epsilon_vs_time}}
\end{figure}

\begin{figure}[hpt]
    \centering
    \includegraphics[width=0.5\textwidth]{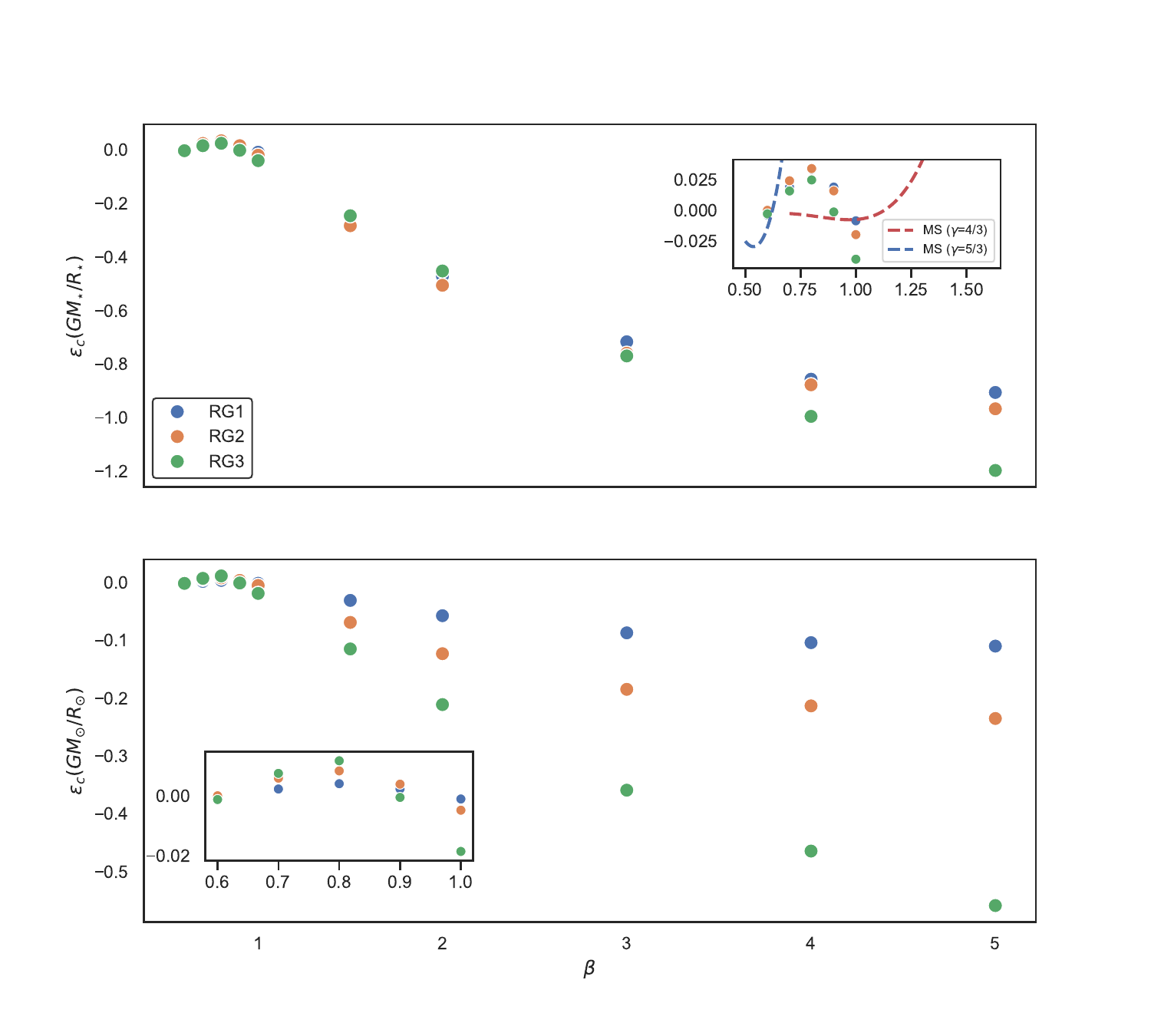}
    \caption{Final specific orbital energy ($\epsilon_c$) with different $\beta$. The upper and lower panel correspond to $\epsilon_c$ normalized to $GM_{\star}/R_{\star}$ and $GM_{\odot}/R_{\odot}$, respectively. The zoomed-in insets show the results of $\beta\lesssim 1$. In the inset in the upper panel, the dashed red and blue lines represent main-sequence star (MS) disruptions for the $\gamma=5/3$ and $\gamma=4/3$ cases studied by \cite{2024ApJ...977...80C}, respectively (see the main text for details). \label{fig: epsilon_vs_beta}}
\end{figure}

To compare the final orbital energies with different $\beta$, we adopted the specific orbital energy averaged over the last $10\,t_{\mathrm{dyn}}$, as illustrated in Fig. \ref{fig: epsilon_vs_beta}. For $\beta\lesssim 0.8$, the remnant core becomes more tightly bound as $\beta$ decreases, whereas it becomes unbound as $\beta$ increases. This characteristic is similar with the previous simulations for the coreless star \citep{2013ApJ...771L..28M,2023MNRAS.520L..38C,2024ApJ...977...80C}.

However, unlike the results for a coreless star where the remnant material becomes more unbound with increasing $\beta$ until it is fully disrupted, the energy change of the remnant core of a giant star shows a reversal evolution around $\beta\sim 0.8$. Beyond this threshold, the orbital energy decreases with increasing $\beta$, entering the negative regime; this indicates that the core remains bound to the SMBH. Moreover, the magnitude of the energy change is significantly larger for $\beta \gtrsim 1$ than for $\beta \lesssim 1$.  

This behavior is significantly different from that of coreless stars, where the change in orbital energy is always positive \citep{2013ApJ...771L..28M,2024ApJ...977...80C}. We adopted the approximate analytical formula for the change in orbital energy of coreless stars from \cite{2024ApJ...977...80C} and  compared the formula results with our simulation results. It should be noted that the analytical formula is only applicable within a limited $\beta$ range, and thus only these regions are plotted in Fig. \ref{fig: epsilon_vs_beta}. It can be observed that as $\beta$ increases, coreless stars gain more energy, which is significantly different from the case of giants.

\subsection{Asymmetric mass loss}

Previous studies of the coreless star indicate that, for small $\beta$, the orbital energy evolution is governed by tidally excited oscillations, whereas for larger $\beta$ it is dominated by asymmetric mass loss, i.e., the difference between mass lost through Lagrangian points $L_1$ and $L_2$ \citep{2013ApJ...771L..28M,2024ApJ...977...80C}. The simulations of \cite{2013ApJ...762...37L} further demonstrate that asymmetric mass loss also dominates the energy variation when a core is present. Below we analyze the mass loss in our simulations as the material that remains unbound to the remnant core.

Material bound to the remnant core is defined as that lying within the Hill sphere, whose radius is $R_{\text{bound}}(t)=\left[M_{\text{bound}}(t)/M_{\text{BH}}\right]^{1/3}r_c(t)$, where $r_c$ denotes the instantaneous separation between the remnant core and the SMBH, and $M_{\text{bound}}(t)$ is the mass enclosed within $R_{\text{bound}}(t)$ \citep{2012ApJ...757..134M}. The bound and unbound masses were tracked as functions of time. Asymmetric mass loss was quantified by comparing the distances of SPH particles from the SMBH with that of the core \citep{2013ApJ...771L..28M}. Mass lost through the Lagrangian points $L_1$ ($\Delta m_1>0$) and $L_2$ ($\Delta m_2>0$) corresponds to gas that becomes unbound from the remnant core and is located closer to and farther from the SMBH, respectively.

\begin{figure}[hpt]
    \centering
    \includegraphics[width=0.5\textwidth]{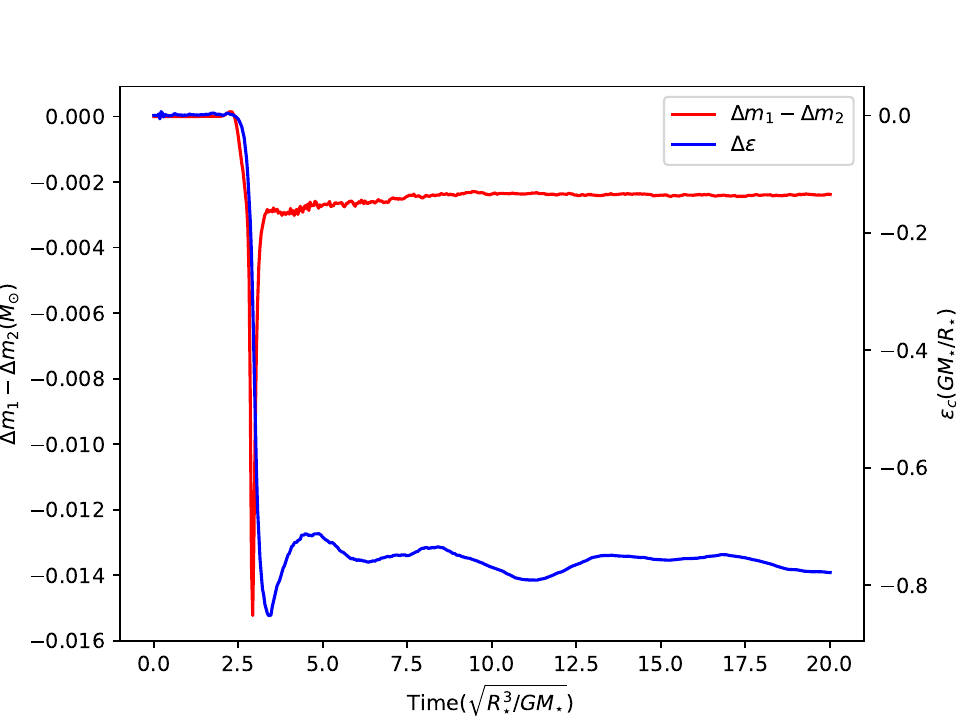}
    \caption{Evolution of the asymmetric mass loss and the specific orbital energy of the remnant material for RG2 with $\beta=3$. The red and blue lines represent the asymmetric mass loss and the specific orbital energy of the remnant core normalized to $GM_{\star}/R_{\star}$, respectively.\label{fig: epsilon_massloss_vs_time}}
\end{figure}

Figure \ref{fig: epsilon_massloss_vs_time} shows the evolution of asymmetric mass loss and the specific orbital energy of remnant material for RG2 with $\beta=3$ as an example. Both asymmetric mass loss and energy variation change dramatically at the moment of disruption and then rebound in the opposite direction. Unlike the subsequent oscillations in energy variation, the asymmetric mass loss stabilizes after the rebound. Therefore, its final value in the simulation can serve as a reference value.

\begin{figure}[hpt]
    \centering
    \includegraphics[width=0.5\textwidth]{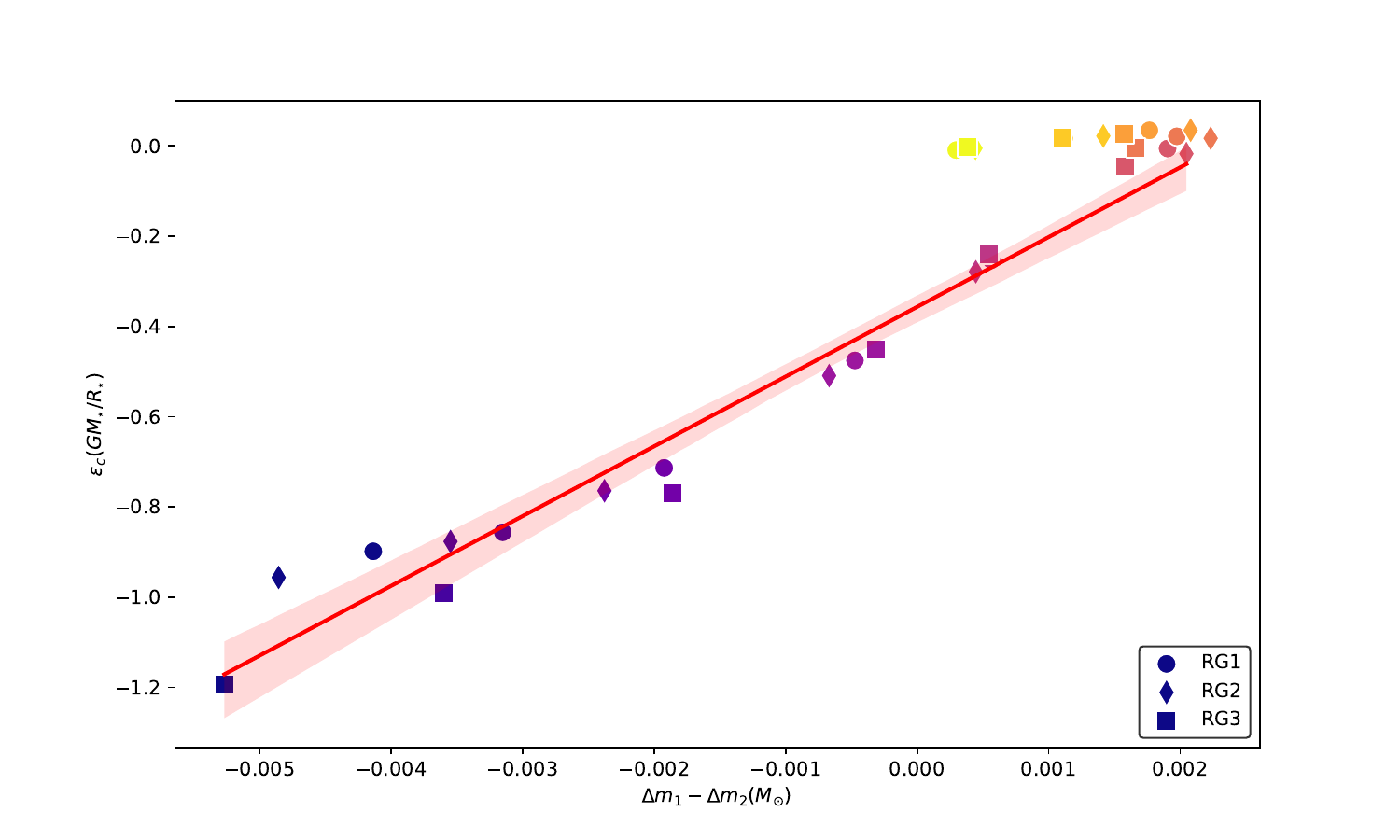}
    \caption{Relationship between the averaged specific orbital energy and the asymmetric mass loss. The symbol colors correspond to different values of $\beta$, matching the scheme in Fig. \ref{fig: epsilon_vs_time}. The solid red line represents the best linear fit for cases where $\beta\le 1$, with the red shaded area indicating the 95\% confidence interval. Note that the averaged specific orbital energy is normalized to $GM_{\star}/R_{\star}$ \label{fig: epsilon_vs_dm}}
\end{figure}

For all simulated cases, Fig. \ref{fig: epsilon_vs_dm} illustrates the relationship between the averaged specific orbital energy and asymmetric mass loss. As $\beta$ increases, $\Delta m_1-\Delta m_2$ initially rises before decreasing to negative values. When this trend reverses, a significant correlation emerges between the energy variation change and asymmetric mass loss, consistent with the findings of \cite{2013ApJ...762...37L}.  

We performed a linear fit to the energy variation with asymmetric mass loss for $\beta\ge 1$,
\begin{equation}
    \epsilon_c = (154.7\pm 7.2) (\Delta m_1-\Delta m_2)-0.357\pm 0.019.
\end{equation}
The correlation coefficient between these quantities is $R^2=0.9669$,  with a p-value of $2.86\times 10^{-13}$. Notably, this correlation only emerges when energy is normalized to $GM_{\star}/R_{\star}$, which can represent the gravitational potential at the surface of the star.

\section{Discussion}

\subsection{Dynamical mechanism}

There are two models that explain the energy variation of the remnant material during partial tidal disruption. The first model proposes that the stripped material interacts with the residual material \citep{2013ApJ...771L..28M}: material lost through the Lagrangian point $L_1$ accelerates the residual material, while material lost through the Lagrangian point $L_2$ decelerates it. Consequently, asymmetric mass loss induces an energy change in the residual material. 

The second model attributes the energy variation to the reformation of the remnant core due to differing density evolution between bound and unbound material \citep{2025MNRAS.542L.110C}. After the star passes periastron, asymmetric density evolution shifts the location of maximum density. As the tidal force from the SMBH weakens during this phase, a remnant core reforms at the new density maximum, altering the orbital energy of the remnant core.

Both models can account for the energy variation of remnant material during the partial tidal disruption of a main-sequence star. However, the case for giant stars differs significantly.

The strong correlation between energy variation and asymmetric mass loss in simulations suggests that the interaction between mass loss and the core plays a dominant role in energy variation. Specifically, when energy is measured in terms of the surface gravitational potential $GM_{\star}/R_{\star}$, simulations with different star structures exhibit a similar correlation, further indicating that the energy variation arises from gravitational interactions between the lost material and the remnant.

However, \cite{2013ApJ...771L..28M} argue that asymmetric mass loss results from the octupole term of the tidal force exerted by the SMBH, which would only enhance mass loss through the $L_1$ point, thereby making the remnant's orbit more unbound. Consequently, this model cannot explain the energy variation of the remnant material in the case of a giant star.

The second model (reformation of the remnant core) is also unsuitable for giant stars, as asymmetric density evolution cannot shift the location of maximum density, since the densest region is always the compact core. Thus, neither model adequately explains the energy variation of remnant material in the partial tidal disruption of a giant star.  

Our simulations strongly support the idea that the energy variation originates from asymmetric mass loss, but the underlying mechanism remains unclear. Two key questions arise:  why mass is preferentially loss through the $L_2$ point rather than the $L_1$ point when $\beta$ is sufficiently large, and how the mass loss interacts with the remaining core. \cite{2013ApJ...771L..28M} only considered instantaneous interactions and attributed the asymmetric mass loss to the octupole term of the tidal force, which evidently cannot answer the first question.  

Regarding the second question, \cite{2024ApJ...977...80C} propose that asymmetric mass loss affects the energy of the remaining core by transferring tidal potential, giving\footnote{\cite{2024ApJ...977...80C} replaced the periastron distance ($R_p$) with the tidal radius ($R_t$) in their formula to align with their simulations. Here we retain the original formulation, so the formula includes an additional factor, $\beta^2$.}  

\begin{equation}
    \epsilon_c \simeq \frac{GM_{\star}}{R_{\star}} \frac{\Delta m_1-\Delta m_2}{M_{rem}}q^{1/3}\beta^2,
\end{equation}
where $M_{rem}$ is the remaining mass of the star and $q=M_{BH}/M_{\star}$ is the mass ratio, approximately $10^6$ in our simulations. \cite{2013ApJ...771L..28M} adopted a different approach, calculating the energy variation via the gravitational force between the disrupted and remaining material, with an interaction time $t_p=\sqrt{R_p^3/G/M_{BH}}$, yielding the specific impulse
\begin{equation}
    I_{1,2}\simeq \frac{G \Delta m_{1,2}}{R_{\star}^2}\sqrt{\frac{R_p^2}{GM_{BH}}}.
\end{equation}
The resulting energy change is
\begin{equation}
    \epsilon_c =\frac{1}{\sqrt{2}}(I_1-I_2)v_p\simeq\frac{GM_{\star}}{R_{\star}}\frac{\Delta m_1-\Delta m_2}{M_{\star}}\frac{q^{1/3}}{\sqrt{2}\beta}.
\end{equation}

\begin{figure}[hpt]
    \centering
    \includegraphics[width=0.5\textwidth]{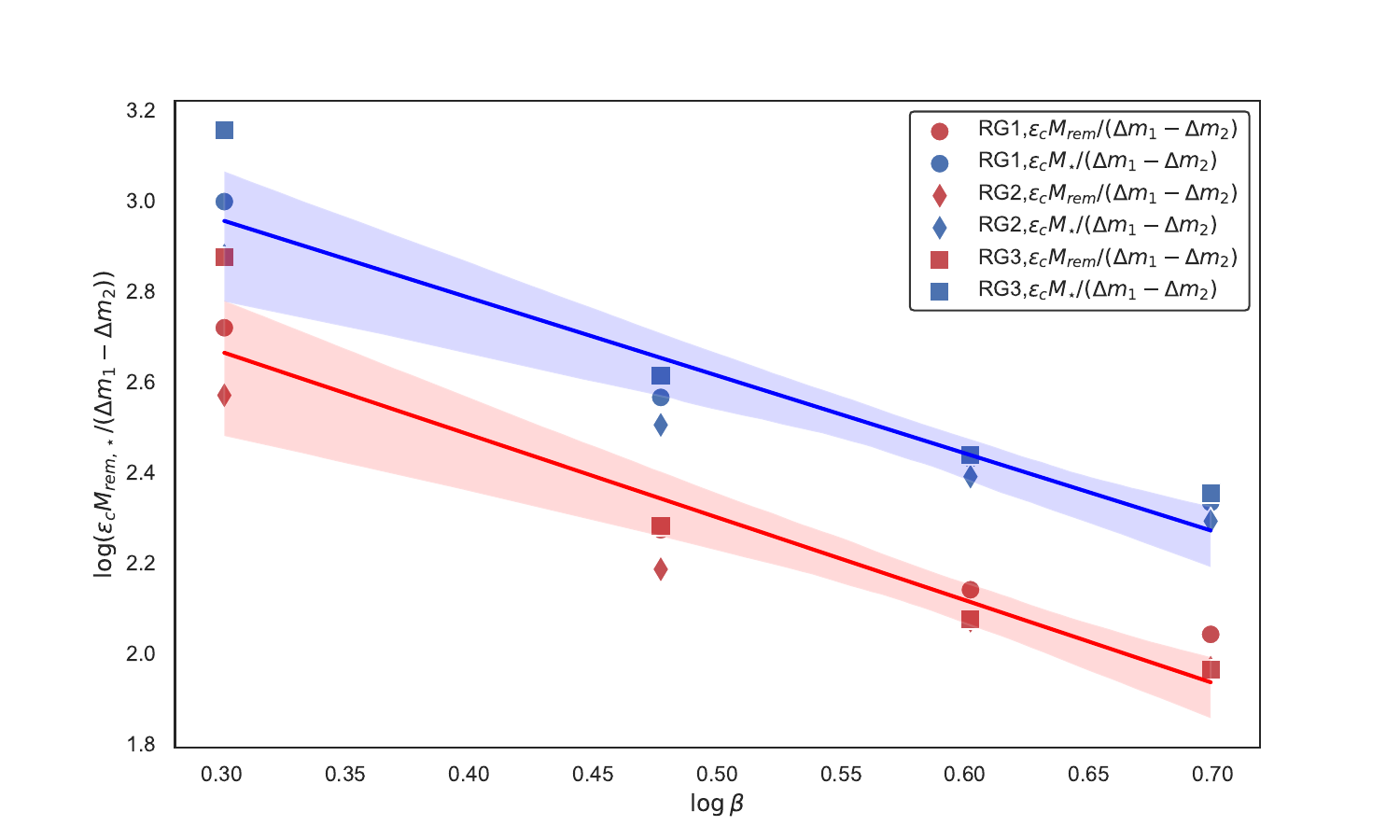}
    \caption{Dependence of $\epsilon_cM_{\star,rem}/(\Delta m_1 -\Delta m_2)$ on $\beta$ in our simulations, where $\epsilon_c$ is normalized to $GM_{\star}/R_{\star}$. The solid blue line and shaded region represent the best-fit linear relationship between $\epsilon_c M_{\star}/(\Delta m_1-\Delta m_2)$ and $\beta$ in logarithmic coordinates and the corresponding 95\% confidence interval, respectively. The solid red line and shaded region show the equivalent relationship for $\epsilon_c M_{rem}/(\Delta m_1-\Delta m_2)$. \label{fig: beta_vs_epsilon_dm}}
\end{figure}

The key distinction between these models lies in their differing dependence of $\epsilon_c$ on $\beta$, which can be tested against our simulation results. Figure \ref{fig: beta_vs_epsilon_dm} displays $\epsilon_c M_{rem}/(\Delta m_1-\Delta m_2)$ and $\epsilon_c M_{\star}/(\Delta m_1-\Delta m_2)$ normalized to $GM_{\star}/R_{\star}$, as a function of $\beta$. The fitting yields  
\begin{equation}
    \epsilon_c M_{rem}/(\Delta m_1-\Delta m_2) = 10^{3.22\pm 0.11}\beta^{-1.83\pm 0.20} 
\end{equation}
and
\begin{equation}
    \epsilon_c M_{\star}/(\Delta m_1-\Delta m_2) = 10^{3.47\pm 0.10}\beta^{-1.72\pm 0.19}.
\end{equation}

These results contradict both models. The energy change exceeds model predictions ($\sim q^{1/3}=10^2$) by an order of magnitude, and the $\beta$ dependence ($\sim \beta^{-1.5\sim -2}$) differs significantly, particularly from the \cite{2024ApJ...977...80C} results. However, their simulations for main-sequence stars align well with their model, suggesting that the core in partial tidal disruption events (pTDEs) may substantially influence the dynamical process.  

This discrepancy suggests that the presence of a dense core significantly alters the mass loss process during partial tidal disruption, differing from scenarios without such a core. A realistic model would need to account for the dynamical evolution of the debris under both the SMBH's tidal force and the self-gravity of the star, which lies beyond the scope of this paper.

\subsection{Repeating partial tidal disrupted events}

After the initial partial tidal disruption, the star can undergo subsequent disruptions upon returning to periastron. Empirical relations indicate that the radius of the giant star depends solely on its core mass \citep{1995MNRAS.273..731R}. Since the core mass remains unchanged during a pTDE, the radius can recover to its original value. This recovery was recently demonstrated by \cite{2025ApJ...983..177N} through MESA simulations, which showed that giant stars regain equilibrium after a thermal timescale, reverting to their initial radius.  

As discussed in \cite{2025ApJ...983..177N}, the nature of repeating pTDEs depends on the orbital period of the remnant star. If the orbital period is not significantly shorter than the thermal timescale, the remnant has sufficient time to recover its original radius. In this scenario, each partial disruption occurs at the same $\beta$ value, progressively stripping the envelope and leaving behind a lower-mass giant star.  

Conversely, if the orbital period is much shorter than the thermal timescale, the star cannot fully restore its radius, and $\beta$ decreases with each subsequent disruption. Although the radius does not return to its initial value, the envelope continues to be stripped until only a low-mass star remains.

A third possibility arises when the orbital period is comparable to the remaining lifetime of the giant star. In this case, the star can evolve into a white dwarf before returning to periastron, thereby terminating the cycle of pTDEs.  

\subsubsection{Orbital period}
The orbital period of the remnant star can be derived from the specific orbital energy,
\begin{equation}\label{eq: orbital period}
    P=2\pi GM_{BH}(-2\epsilon)^{-3/2}\approx 112\frac{M_{BH}}{10^6\,M_{\odot}}\left(\eta\frac{M_{\star}}{M_{\odot}}\frac{R_{\odot}}{R_{\star}}\right)^{-3/2}\,\mathrm{years},
\end{equation}
where $\eta=\epsilon/(GM_{\star}/R_{\star})$ represents the extracted fraction of the specific self-bound energy of the captured star. Our simulations show that $\eta$ typically ranges between $0.2$ and $1.2$ for $\beta\gtrsim 1$.  

The Hills mechanism can also produce a repeatedly disrupted bound star \citep{2022ApJ...929L..20C}. The corresponding specific orbital energy of the captured star can be estimated from the tidal potential,
\begin{equation}\label{eq: epsilon_H}
    \epsilon_H=-\frac{GM_{BH}}{R_H^2}a_b=-\frac{GM_b}{a_b}\left(\frac{M_{BH}}{M_b}\right)^{1/3},
\end{equation}
where $R_H=(M_{BH}/M_b)^{1/3}a_b$ is the Hills disruption radius, with $a_b$ and $M_b$ denoting the semimajor axis and mass of the binary star, respectively.  

For a disrupted binary with a similar mass and scale to a single giant star, Eq. \eqref{eq: epsilon_H} demonstrates that the Hills mechanism yields a significantly lower specific orbital energy, which corresponds to shorter orbital periods. For a mass ratio $M_{BH}/M_b=10^6$, the specific orbital energy generated via the Hills mechanism is $10^2$ times lower and the orbital period is $10^3$ times shorter than those in tidal capture, consistent with the discussion in \cite{2022ApJ...929L..20C}.  

The period of rpTDEs formed through the Hills mechanism is substantially shorter than that of rpTDEs produced by tidal capture. Consequently, tidal capture can only explain observed rpTDEs with long periods, which also require low-mass SMBHs and early evolved stars. On the other hand, since tidal capture inevitably occurs during deep disruption, this mechanism can generate numerous long-period rpTDEs. Such rpTDEs can resemble "standard" TDEs \citep{2025ApJ...987L..20M}, suggesting that the tidal capture of giant stars could enhance the TDE rate and form lighter giants \citep{2025ApJ...983..177N}.  

When $\beta$ is small, the star gains a positive energy, corresponding to a kick velocity defined by \cite{2013ApJ...771L..28M} as\begin{equation}
    v_{\text{kick}}=\sqrt{2\epsilon_c}\approx 62\sqrt{\frac{\eta}{0.01}\frac{M_{\star}}{M_{\odot}}\frac{R_{\odot}}{R_{\star}}} \,\mathrm{km/s}.
\end{equation}
When a giant is kicked, the extract fraction ($\eta$) is much smaller than that of the main-sequence star, which can reach a maximum of a unit, meaning the kick velocity is an order of magnitude lower.

\subsubsection{Possible sources}

In principle, all rpTDEs with very long periods can be explained by the tidal capture of a giant star. Among the published rpTDEs identified in optical and UV bands, only F01004-2237 and RX J133157.6-324319.7 exhibit periods exceeding $\sim 10$ years \citep{2025ApJ...985..175S,2024A&A...692A.262S,2022RAA....22e5004H,2023MNRAS.520.3549M}. However, their SMBH masses, estimated at $10^{6.5} M_{\odot}$ and $10^{7.4} M_{\odot}$, respectively \citep{2025ApJ...985..175S}, are too high to produce such long periods according to Eq. \ref{eq: orbital period}. Thus, the Hills mechanism is favored to explain these rpTDEs \citep{2022ApJ...929L..20C}.

Several potential repeating transient sources, such as GSN 069, may also be rpTDEs. Initially detected as a large X-ray flux increase followed by a long-term decay, GSN 069 was proposed to be either a TDE or a highly variable active galactic nucleus (\citealt{2013MNRAS.433.1764M,2018ApJ...857L..16S}). Subsequent observations of anomalous carbon-to-nitrogen abundance ratios provided evidence of a pTDE involving an evolved star \citep{2021ApJ...920L..25S}.  

GSN 069 is also the first galaxy discovered to exhibit quasi-periodic eruptions (QPEs), which emerged during the late stages of its TDE \citep{2019Natur.573..381M}. While the origin of QPEs remains uncertain, they are thought to arise from objects orbiting the SMBH either colliding with the accretion disk \citep{2023ApJ...957...34L} or being accreted by the SMBH \citep{2019ApJ...886L..22W,2022ApJ...933..225W,2022A&A...661A..55Z,2020MNRAS.493L.120K}. Following the appearance of QPEs, a second TDE flare occurred in GSN 069, suggesting an rpTDE with an interval of approximately 10 years \citep{2023A&A...670A..93M}.  

The second flare implies the presence of additional material falling back toward the SMBH. \cite{2024A&A...682L..14W} propose that a TDE involving a differentially rotating envelope—such as a common envelope—could produce a second flare. However, this model can generate at most two additional flares. If further flares are observed, GSN 069 would instead represent the repeating disruption of a bound star in an orbit with a period of roughly 10 years.  

The SMBH mass in GSN 069 is estimated to be $0.4-1 \times 10^{6} M_{\odot}$ \citep{2023A&A...670A..93M}. If the disrupted star is a low-mass giant with a high impact parameter ($\beta$; where $\eta$ can exceed unity), the post-TDE orbital period could reach 10 years, as inferred from Eq. \ref{eq: orbital period}. This scenario could explain the second flare in GSN 069 and predicts additional flares in the future.  

In order to produce rpTDEs with shorter periods, low-mass SMBHs and compact disrupted stars are required. TDEs involving IMBHs and white dwarfs could produce periods that are short enough to be detectable. Although our study focuses on the pTDEs of giant stars, white dwarfs—despite their thin envelopes—also possess dense cores, making them plausible candidates for tidal capture.  

For instance, a $1 M_{\odot}$ white dwarf with a radius of $0.01 R_{\odot}$ disrupted by a $10^5 M_{\odot}$ IMBH could be bound in an orbit of $\sim 40$ days ($\eta=1$). However, no confirmed cases of white dwarfs disrupted by IMBHs have been observed to date, as the observational signatures of such events remain unclear. The repetitive nature of tidal capture could provide constraints for identifying these events.

\subsection{Impact on the orbital evolution}  

When a star enters a bound orbit, the instantaneous change in orbital energy during each pTDE influences the orbital evolution. This short-term effect was ignored in previous studies that examined the impact of mass transfer on highly eccentric orbits around an SMBH; they focused solely on long-term effects caused by mass changes \citep{2020MNRAS.493L.120K,2024A&A...687A.295W,2025MNRAS.543.3503Y}.

The effect of mass transfer depends on the direction of mass and angular momentum transfer. The change in total angular momentum is given as
\begin{equation}
    \frac{\dot{J}}{J}=\frac{\dot{M}_{BH}}{M_{BH}}+\frac{\dot{M}_{\star}}{M_{\star}}-\frac{\dot{M}_{t}}{2 M_{t}}+\frac{1}{2} \frac{\dot{e}}{1+e}+\frac{1}{2} \frac{\dot{r}_{p}}{r_{p}},
\end{equation}
where $M_t=M_{\star}+M_{BH}$ is the total mass. Assuming mass transfer occurs instantaneously at periastron ($\dot{r}_p=0$), we obtain \citep{2024A&A...687A.295W}
\begin{equation}
    \frac{\dot{a}}{a}=-\mathcal{F}(\gamma,q,\lambda)\frac{1 + e}{1 - e}\frac{\dot{M}_{\star}}{M_{\star}},
\end{equation}
where
\begin{equation}
    \mathcal{F}(\gamma,q,\lambda)=\frac{2 + 2(\gamma - 1)\lambda+(1 - \gamma)q-2\gamma q^{2}}{1 + q}.
\end{equation}
Here $\gamma=-\dot{M}_{BH}/\dot{M}_{\star}$ represents the fraction of stellar mass that is lost and accreted by the SMBH, and $\lambda$  is the ratio of the specific angular momentum carried away by mass loss to the specific angular momentum of the star, such that $\dot{J}/J=\lambda \dot{M}_t/(qM_t)$. The change in specific orbital energy per mass transfer event is

\begin{equation}
\begin{aligned}
    \Delta E_{MT}&=\frac{GM_t}{a}\left(\frac{\dot{a}}{a}-\frac{\dot{M}_t}{M_t}\right)\\
    &\approx -\frac{4M_t}{r_p}\frac{\Delta M_{\star}}{M_{\star}}[1+(\gamma-1)(\lambda+q)]\\
    &\approx -4\beta q^{-2/3}\frac{\Delta M_{\star}}{M_{\star}}[1+(\gamma-1)(\lambda+q)]\frac{GM_{\star}}{R_{\star}},
\end{aligned}
\end{equation}

\noindent where we ignore the $q^2$ term and assume $e\approx 1$. Notably, mass transfer increases orbital energy, counteracting the energy decrease during a partial disruption at high $\beta$. The ratio of these competing effects is

\begin{equation}\label{eq: effect ratio}
    \left|\frac{\Delta E_{MT}M_{BH}M_{\star}/M_t}{\Delta\epsilon_c M_{\star}}\right|\approx 4\frac{\beta q^{-2/3}}{\eta}\frac{\Delta M_{\star}}{M_{\star}}[1+(\gamma-1)(\lambda+q)].
\end{equation}

Clearly, the dominant effect depends on the mass ratio and mass loss, scaling as $\sim q^{-2/3}\Delta M_{\star}/M_{\star}$. For large $\beta$, $\Delta M_{\star}/M_{\star} \gtrsim 0.1$, rendering the instantaneous effect negligible. However, for $\beta \approx 0.5$, mass loss is minimal, and the instantaneous effect can surpass mass transfer effects. In this regime, tidal oscillations dominate and reduce the orbital energy, requiring analysis via traditional tidal dissipation theory \citep{2014ARA&A..52..171O}.  

The above analysis assumes that the $q$ term can be neglected in Eq. \ref{eq: effect ratio}. However, if $\lambda(1-\gamma)=1$, meaning all angular momentum carried away by mass loss from the star is lost (even when some mass is accreted by the SMBH), the situation changes. This scenario occurs because matter accreted by the SMBH through the inner accretion disk carries minimal angular momentum. Most angular momentum is transferred outward through the disk and not returned to the orbit. In this case, the ratio of the two effects becomes $\sim q^{1/3}\Delta M_{\star}/M_{\star}$, causing the instantaneous effect to dominate as the impact of mass change is substantially reduced.  

Overall, the instantaneous effects of pTDE on orbital evolution are generally negligible. These effects become significant only when the angular momentum lost through mass loss is entirely removed from the orbit. However, this condition is unlikely to persist for an extended duration, as the expanding accretion disk will tidally interact with the star, transferring angular momentum back to the orbit. Consequently, pTDEs may dominate orbital evolution solely during the initial phase of rpTDEs, when the accretion disk has just formed. This hypothesis could potentially be verified through future long-term observations of confirmed sources, where the direction of the orbital period would reverse, shifting from contraction to expansion.

\section{Conclusion}

We performed hydrodynamic simulations of the pTDEs of giant stars and found that the dynamical evolution of remnant material during this process is significantly different compared to main-sequence stars. When $\beta$ is small, the remnant material from the partially disrupted giant star behaves similarly to that of a main-sequence star, gaining a kick velocity and becoming more unbound in its orbit. However, the giant star can survive at larger $\beta$, in which case the remnant material after disruption remains bound to the SMBH.

Although the outcomes of dynamical evolution differ, the changes in orbital energy after the disruption of giant stars are still strongly correlated with the asymmetric mass loss, similar to main-sequence stars. This suggests they share a similar evolutionary mechanism, where the variation in orbital energy arises from the difference in interaction between the remnant material and the mass lost through the Lagrange points $L_1$ and $L_2$. And it rules out the model by \cite{2025MNRAS.542L.110C}, according to which the energy change arises from the drift of the maximum density position due to the evolution of remnant matter density, followed by the reformation of the remnant core at the density maximum. 

Nevertheless, the magnitude of the orbital energy changes in the simulations and its dependence on $\beta$ still contradict the predictions of the both current models,  as they yield smaller energy variations and a more gradual dependence on $\beta$. Given that the current model only roughly accounts for the gravitational potential between the remnant and lost matter or their instantaneous interactions during disruption \citep{2024ApJ...977...80C,2013ApJ...771L..28M}, the discrepancy with the simulation results implies the need to consider longer-term interactions between the remnant and lost matter, where the presence of a dense core may play a non-negligible role.

On the other hand, our simulations also indicate that the giant star is inevitably captured by the SMBH after a deep pTDE, subsequently inducing rpTDEs. However, since the specific orbital energy generated by tidal capture is comparable to the specific self-binding energy of the giant star, it will be significantly lower than the specific orbital energy produced by the Hills mechanism, which is equivalent to the tidal potential experienced by the binary \citep{2022ApJ...929L..20C}. Tidal capture can only produce rpTDEs with periods of hundreds of years or even longer, making it impossible to observe multiple outbursts. Nevertheless, they can be identified as multiple single TDEs, thus contributing significantly to the TDE event rate.

If the SMBH has a very low mass or the disrupted object is extremely dense, the rpTDEs produced by tidal capture can be observed with multiple outbursts. GSN 069 is an rpTDE with a 10 year interval, where the SMBH has a relatively low mass and the disrupted object is a giant star \citep{2023A&A...670A..93M,2021ApJ...920L..25S}, and thus GSN 069 likely formed through tidal capture. Additionally, if the pTDE process of a white dwarf exhibits properties similar to those of a giant star, its higher self-binding energy could produce orbital periods of several tens of days, serving as a constraint for future confirmation of white dwarf TDEs.

The energy changes during the pTDE process can also affect the orbital evolution of accretion systems with high eccentricity and extreme mass ratios. However, this instantaneous effect is significantly weaker than the influence of mass changes on the orbit, unless $\beta$ is very close to 0.5. In such cases, the energy changes are dominated by tidal oscillations, and the orbital evolution can be described by classical tidal theory. There is another scenario where this instantaneous effect can dominate orbital evolution, which occurs when the angular momentum carried by the stripped material is entirely stored in the accretion disk and not returned to the orbit. However, this situation cannot persist for long because the accretion disk eventually transfers the angular momentum back to the orbit through tidal interactions. Thus, it only occurs in the early stages of rpTDEs formed by tidal capture, which can be tested using future long-term observations of confirmed sources.

\begin{acknowledgements}
This work is supported by the National Natural Science Foundation of China (grant Nos. 12494575, 12273009 and 12503051).
\end{acknowledgements}

\bibliographystyle{aa}
\bibliography{ref} 

\appendix
\section{Specific orbital energy determined via different methods}
\label{sec: energy in two ways}
Our simulations use a sink particle to replace the compact core of the giant star, then treat this sink particle as the center of the remnant material. However, previous works use the physical quality of the SPH particles to calculate the velocity and position of the center of the mass (COM)\citep{2013ApJ...771L..28M,2023MNRAS.520L..38C}. 

Following the method in \cite{2023MNRAS.520L..38C}, we calculate the averaged position and velocity of the SPH particles the density of that is greater than 1 per cent of the maximum density, which is treated as the remnant core. Then the specific orbital energy can be obtained by the COM of those bound SPH particles, which can be compared with the result calculated by the sink particle.

\begin{figure}[hpt]
    \centering
    \includegraphics[width=0.5\textwidth]{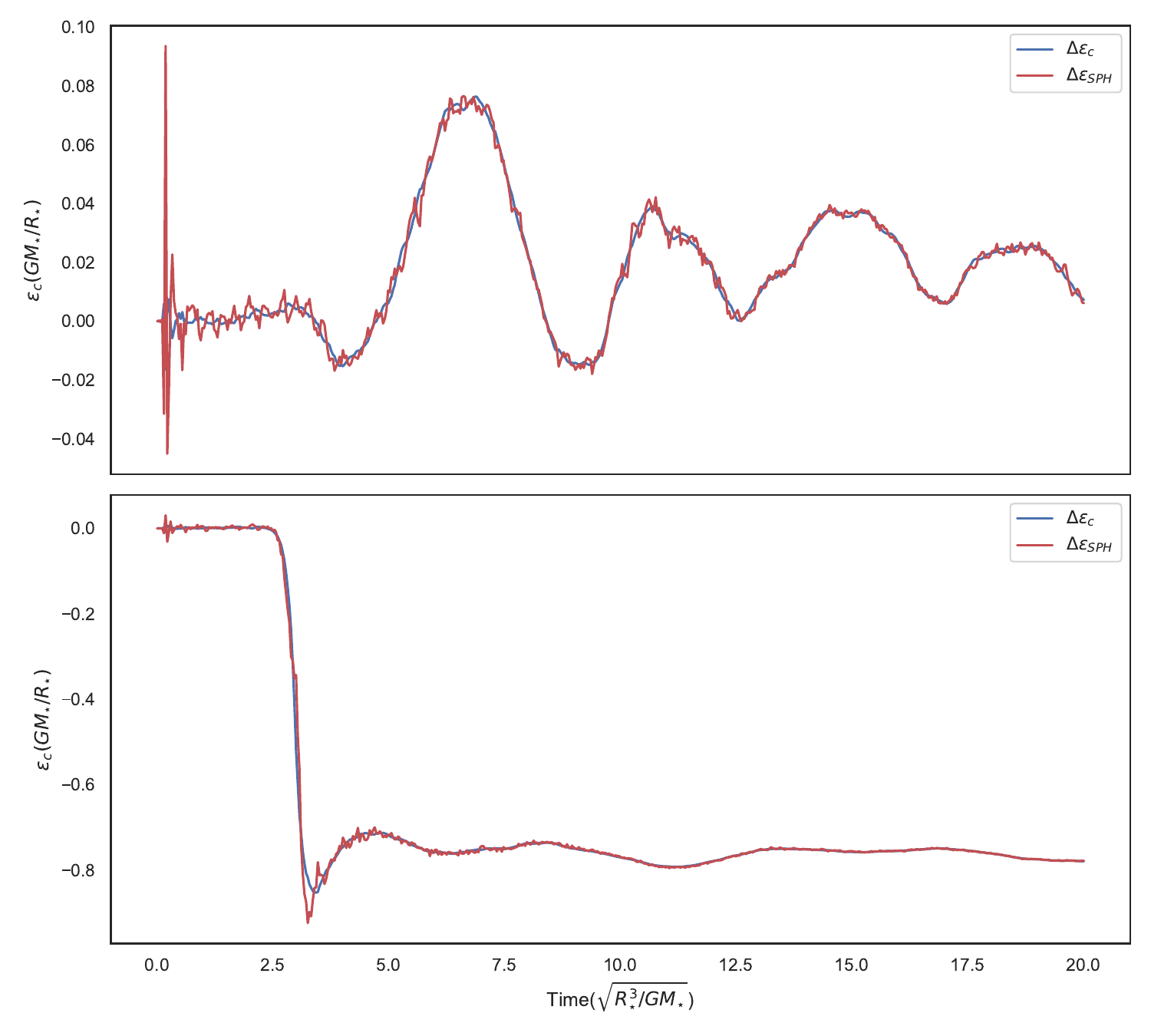}
    \caption{Relation between the averaged specific orbital energy and the asymmetric mass loss determined via two different approaches. The blue lines indicate that the specific energy was calculated using the sink particle, and red lines using the COM of bound SPH particles. Upper panel: Case of RG2 with $\beta=0.7.$ Lower panel: Case of RG2 with $\beta=3$.\label{fig: energy in two ways}}
\end{figure}

In Fig. \ref{fig: energy in two ways} we show the specific orbital energy determined via two approaches for RG2 with $\beta=0.7, 3$ as an example. The specific energies calculated using the two approaches are consistent with each other; however, the value calculated using the sink particle has less noise, indicating that it is more suitable.

\section{Stellar structure in MESA and PHANTOM}
\label{sec: stellar structure comparison}
\begin{figure}[hpt]
    \centering
    \includegraphics[width=0.5\textwidth]{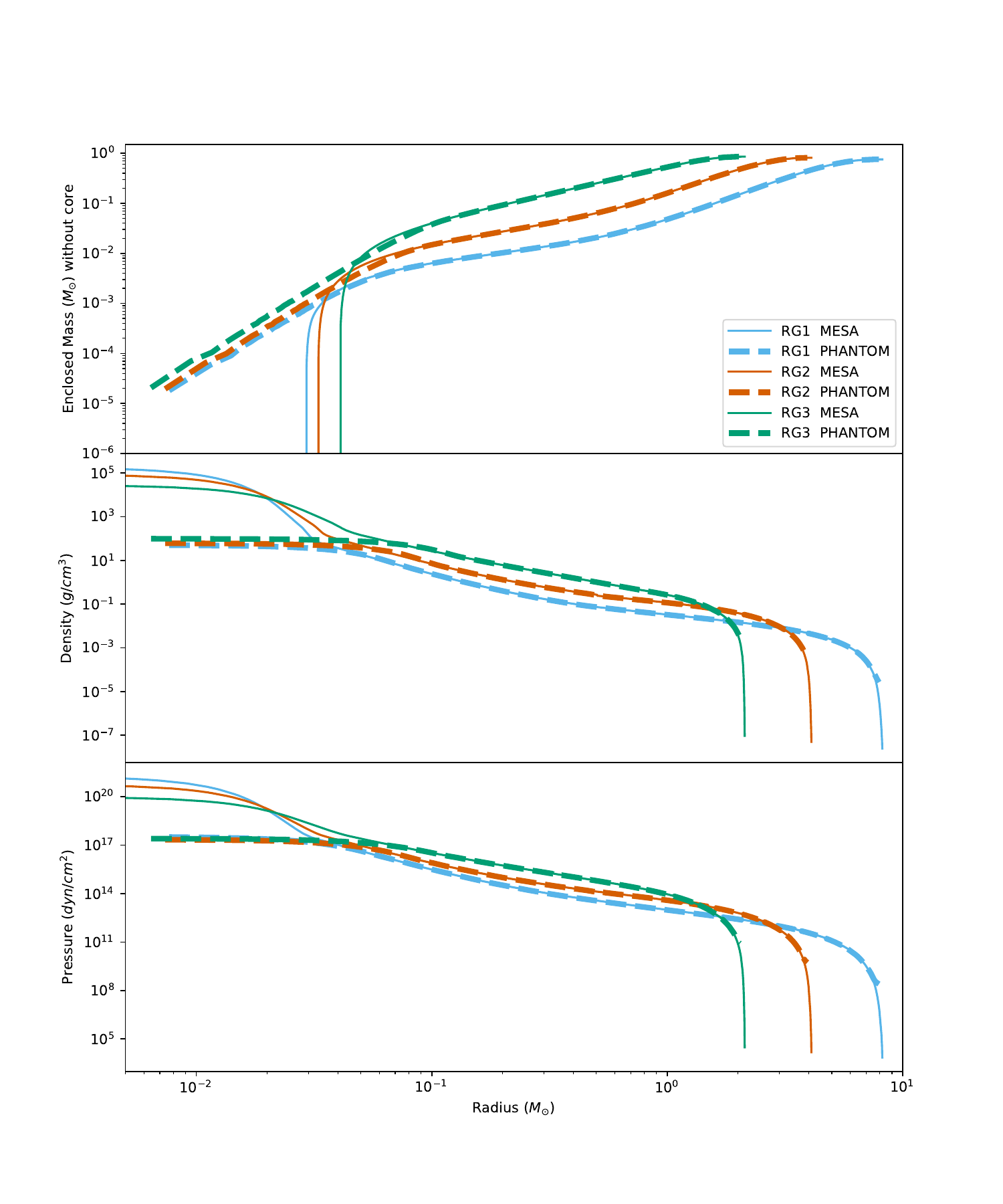}
    \caption{Comparison of the stellar structure of the giant stars in MESA and the relaxed star in PHANTOM. The upper, middle, and lower panels represent the enclosed mass, density, and pressure at different radii, respectively. Blue, red, and green lines represent RG1, RG2, and RG3, respectively. Dashed and solid lines indicate stars in MESA and PHANTOM, respectively. Since the core is replaced by a sink particle, there is a difference at low radius. \label{fig: stellar structure comparison}}
\end{figure}

In our simulation, the initial stellar structure was mapped from the profile obtained with MESA, and subsequently the star was relaxed to achieve hydrostatic equilibrium. We employed a sink particle to represent the compact core of the giant star, while the remaining envelope is modeled as an adiabatic gas. These assumptions may lead to changes in the stellar structure, thus we compare the stellar structure of the star in MESA and the relaxed star in PHANTOM in Fig. \ref{fig: stellar structure comparison}.

It can be seen that the stellar structures in MESA and PHANTOM are almost identical, except that the radius range in PHANTOM is slightly smaller than that in MESA\footnote{It should be noted that the lines of PHANTOM in Fig. \ref{fig: stellar structure comparison} are the results after binning every 50 SPH particles, so the actual radius range is slightly larger than shown in the figure but still smaller than that of MESA.}. The consistency of density and pressure also indicates that using the adiabatic equation of state with an adiabatic index $\gamma=5/3$ to describe the envelope is appropriate.

\section{Mass loss in simulations}

The mass loss in our simulation is obtained by calculating the material outside a Hill sphere with radius 

\begin{equation}\label{eq: hills_1}
    [M_{bound}(t)/M_{BH}]^{1/3}r(t).
\end{equation}
The radius differs from the definition 
\begin{equation}\label{eq: hills_2}
    (M_c/M_{BH})^{1/3}r(t)
\end{equation}
used by \citep{2012ApJ...757..134M}, where they considered the core mass $M_c$ to dominate the residual material mass and thus neglect the gravitational influence of the surrounding material.

\begin{figure}[hpt]
    \centering
    \includegraphics[width=0.5\textwidth]{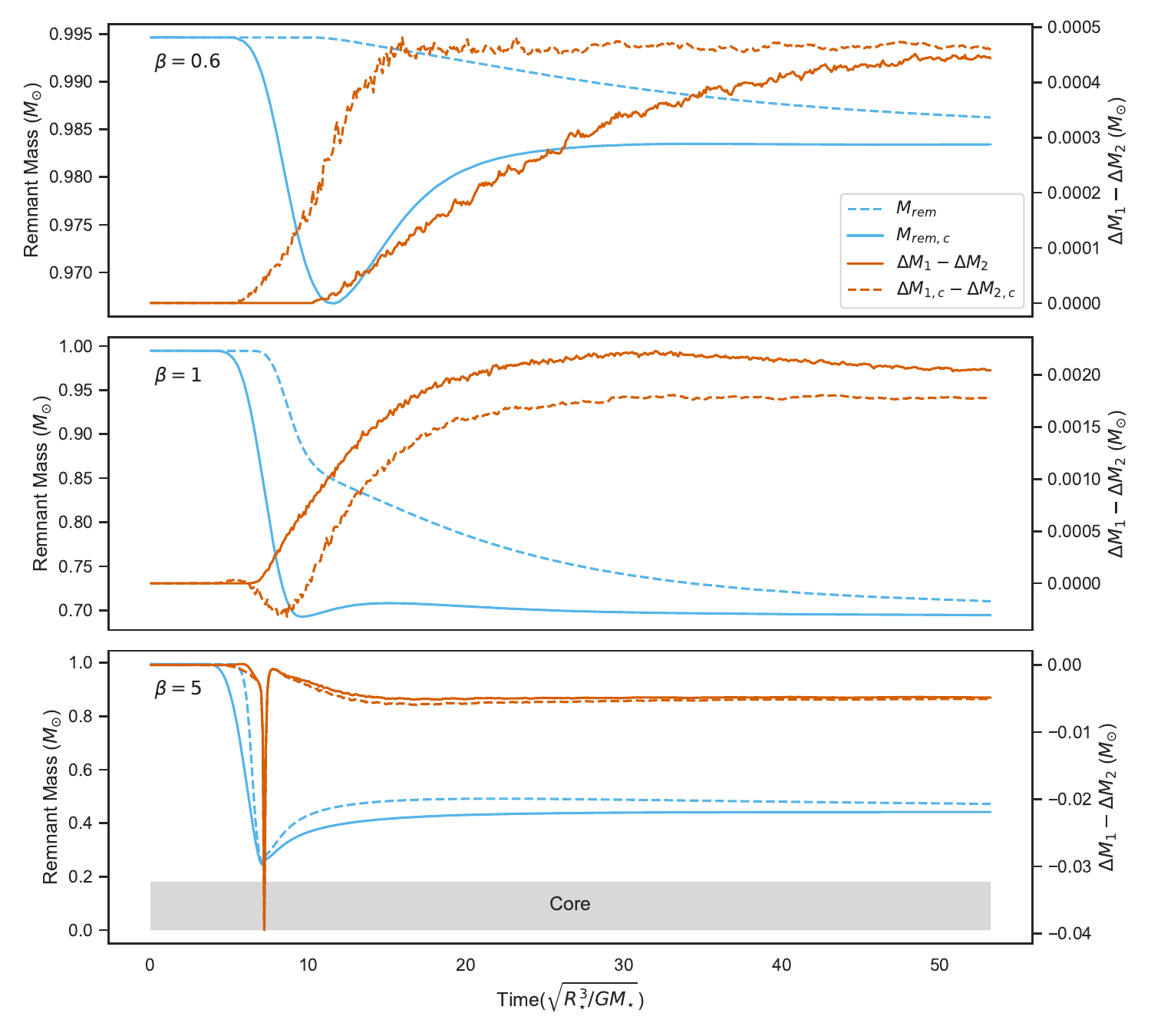}
    \caption{Comparison of the mass loss calculated via two different methods in our simulations. The top, middle, and bottom panels show the simulations of RG2 at $\beta$=0.6, 1, and 5, respectively. The solid and dashed lines represent the two different definitions of the Hill radius from Eqs. \ref{eq: hills_1} and \ref{eq: hills_2}. The gray area in the lower panel shows the dense core. \label{fig: mass loss comparison}}
\end{figure}

As shown in Fig. \ref{fig: mass loss comparison}, the difference between the mass loss curves from the two Hill radius definitions increases as $\beta$ decreases. This is because at smaller $\beta$ values, more material remains around the dense core in the final remnant, resulting in a larger remnant mass when this material is included. Also, the remnant mass from \cite{2012ApJ...757..134M}'s method stabilizes quickly after passing periastron, matching their results, while our method shows later stabilization. The similar difference appears for asymmetric mass loss. However, the overall difference in mass loss between the two methods is small.

\begin{figure}[hpt]
    \centering
    \includegraphics[width=0.5\textwidth]{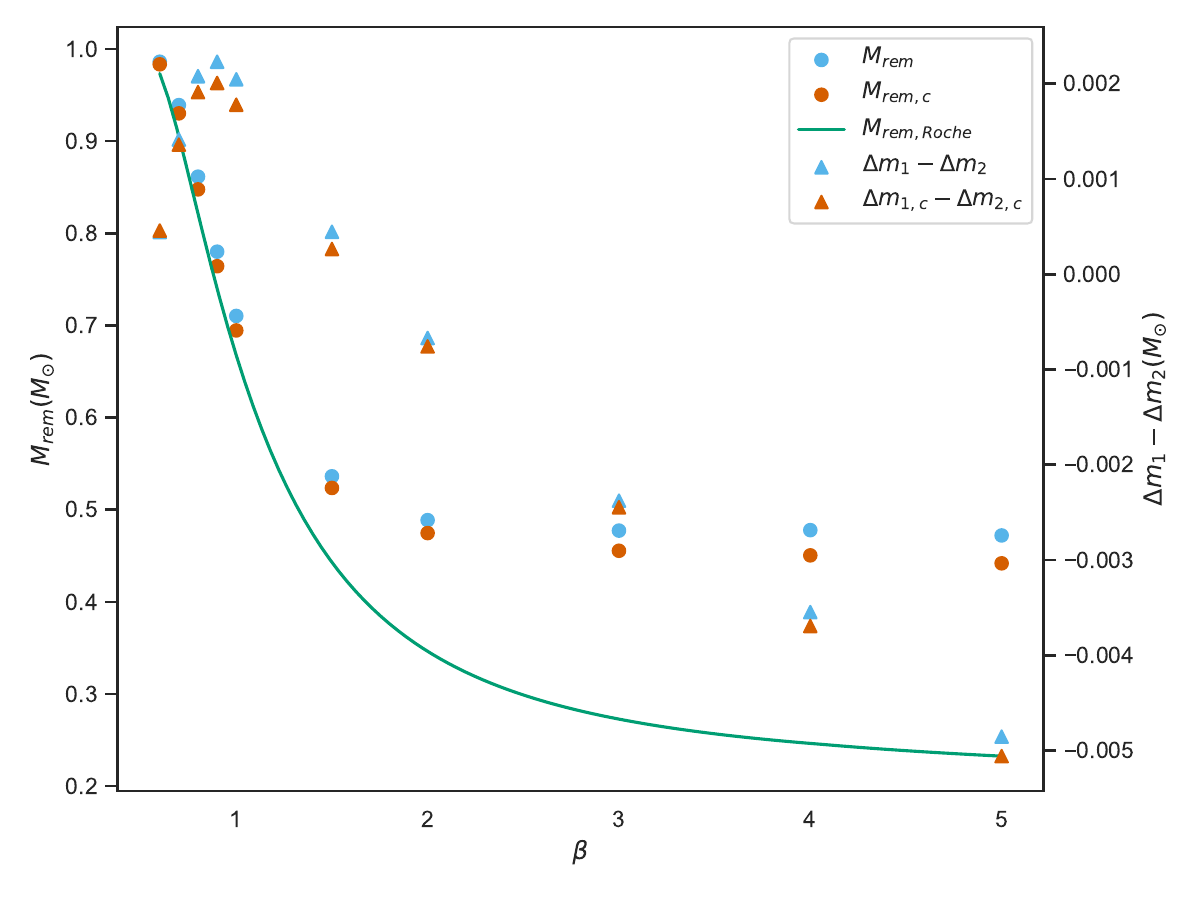}
    \caption{Mass loss for RG2 at different $\beta$ values. Dots and triangles show the remnant mass and asymmetric mass loss, respectively, with blue and red corresponding to the two Hill radius definitions from Eqs. \ref{eq: hills_1} and \ref{eq: hills_2}. The solid green line assumes only material outside the Roche lobe is stripped at periastron. \label{fig: mass loss vs beta}}
\end{figure}

Thus, the relationship between remnant mass, asymmetric mass loss, and $\beta$ is not significantly affected by the definition of the Hills sphere, as seen in Fig. \ref{fig: mass loss vs beta}. When $\beta$ increases, the asymmetric mass transfer first grows and then keeps decreasing, while the reduction in remnant mass slows down until it becomes constant. At low $\beta$, the remnant mass can be obtained by calculating the mass within the Roche lobe at periastron. This approach has also been verified for main-sequence stars \citep{2024ApJ...977...80C} and white dwarfs \citep{2023ApJ...947...32C}. As shown in Fig. \ref{fig: mass loss vs beta}, the simulation results match those from this method at low $\beta$.

\end{document}